\documentclass[final,5p,times,twocolumn]{elsarticle}

\usepackage{graphicx}
\graphicspath{{./img/}}

\usepackage{amssymb}
\usepackage{amsmath}
\usepackage{booktabs}
\usepackage	[
		pdfpagelabels,
		bookmarksnumbered,%
		bookmarksopen,%
        colorlinks=true,%
        citecolor=grey%
        ]{hyperref}
\usepackage{microtype}
\usepackage{framed} 
\usepackage{multicol} 

\usepackage{nomencl} 
\makenomenclature
\renewcommand*\nompreamble{\begin{multicols}{2}}
\renewcommand*\nompostamble{\end{multicols}}

\biboptions{square,numbers}

\journal{Energy and Buildings}

\begin{document}

\begin{frontmatter}



\title{Integrated canopy, building energy and radiosity model for 3D urban design}


\author[EIVP]{Etienne Burdet\corref{cor1}}
\ead{etienne.burdet@eivp-paris.fr}

\author[EIVP]{Morgane Colombert}
\author[UPEMLV]{Denis Morand}
\author[EIVP,UPEMLV]{Youssef Diab}

\cortext[cor1]{Corresponding author}

\address[EIVP]{University Paris-Est, Lab'Urba (EA 3482), EIVP, Paris, France}
\address[UPEMLV]{University Paris-Est, Lab'Urba (EA 3482), UPEM, Champs-sur-Marnes, France}

\begin{abstract}
We present an integrated, three dimensional, model of urban canopy, building energy and radiosity, for early stage urban designs and test it on four urban morphologies. All sub-models share a common descriptions of the urban morphology, similar to 3D urban design master plans and have simple parameters. The canopy model is a multilayer model, with a new discrete layer approach that does not rely on simplified geometry such as canyon or regular arrays. The building energy model is a simplified RC equivalent model, with no hypotheses on internal zoning or wall composition. We use the CitySim software for the radiosity model. We study the effects of convexity, the number of buildings and building height, at constant density and thermal characteristics. Our results suggest that careful three dimensional morphology design can reduce heat demand by a factor of 2, especially by improving insolation of lower levels. The most energy efficient morphology in our simulations has both the highest surface/volume ratio and the biggest impact on the urban climate.
\end{abstract}


\end{frontmatter}


\section{Introduction}

From narrow courtyards and passages in medina quarters or small compact plots in Japanes cities, to large avenues in Paris and terraced band housing in northern Europe, one can only be fascinated by the number and complexity of solutions that traditional urban fabrics present in terms of adaptation to the climate (see for example \citep{Firley2009} for a broad picture of this morphological diversity). All such urban fabrics have in common a very intricate -- and rarely intuitive -- relationship between indoor and outdoor spaces, to find subtle compromises between  access to the sun, compactness and air circulation. With the current world energy situation and the ever growing importance of rapid urbanization, it would be only natural for urban planners and architects to try to design such complex solutions to make urban developments more energy efficient. The critical point in so doing is to go beyond the simple sum of building consumption at an early stage. An urban designer needs to take into account the interactions between buildings and open spaces by considering sun obstruction and the urban heat island effect. Oke \citep{Oke1982} developed the concept of the canyon to explain how  streets trap solar radiations and collect anthropogenic heat released by buildings, resulting in much higher temperatures in urban areas than in their rural counterparts. The canyon concept was later settled into a more physically sound model \citep{Masson2000}, followed by multiple models adding more detailled geometries \citep{Rasheed2011}, turbulent transfers \citep{Martilli2002,Hamdi2008} and Building Energy Models (BEM) \citep{Bueno2012b,Salamanca2009a,Ihara2008a}. The need for BEMs in urban canopy models reveals that building energy and the urban micro-climate are in fact a joint problem. However, up to the best of our current knowledge all urban canopy models\footnote{We refer here to \emph{Urban Canopy Models} as a specific category of models used to compute urban temperatures and turbulent transfers. Other methods may include CFD or zonal methods for example and are not restricted to simplified geometries} reason on simplified geometries -- either canyon or regular arrays \citep{Grimmond2011}. Simplified geometries can prove limited in terms of solar radiation modeling. Three-dimensional morphology can produce complex and counter intuitive irradiation patterns, in consequence greatly impacting building energy demand \citep{Robinson2006,Ratti2003}. Moreover, an urban designer using a canopy model would then have to think about the real geometry for design but about a simplified one for physics, thus greatly reducing the potential of the model in terms of intuition and reasoning. Building on existing models of urban physics, we present an integrated model of urban canopy, BEM and radiative transfers, with a description of the urban morphology as close as possible to three dimensional urban design master plans, and then use it on simple morphological variations.

	\begin{table*}      
      \begin{framed}
		\printnomenclature[2.5em]
      \end{framed}
   \end{table*}   
\section{Model development}
\label{sec:eqs}

After a bibliographic review of indoor and outdoor thermal models responding to an urban or neighborhood scale, we selected a canopy, a BEM and a radiation model that were the closest to the desired level of detail. We then adapted the models so that they all shared the same representation of urban morphology and the same parameters. The parameters need to be as simple as possible and should not require details unknown to urban design stages (e.g. window positioning or internal zoning). To meet such requirements, we first relaxed the regular array hypothesis in the multilayer model of Kondo \citep{Kondo2005}. We next adapted a single zone 2RC BEM with simple parameters. We present a final workflow to couple the analytical model to the CitySim\footnote{Developed at LESO-PB EPFL http://citysim.epfl.ch/} software to compute the radiative net flux.
\subsection{Canopy Model}
\subsubsection{Discrete layer model}
The initial choice for the canopy model was the multilayer model of Kondo \citep{Kondo2005} because it has the most detailed description of urban morphology (regular arrays with height density probability) and can be coupled with a BEM \citep{Ihara2008a}. Our goal in this section was to reformulate the model, but with general floor and wall areas defined storey by storey (or layer per layer), so that the parameters of the equations were actual design parameters. The core idea we used to relax the regular array hypothesis was to apply a spatial discretization corresponding to that of the BEM and radiosity models in the vertical direction.

Multilayer canopy models consider only the vertical diffusion of heat and average it over the horizontal plane. The vertical turbulent transport is derived from the potential temperature, using K-theory $\overline{w\theta_c} = - C_p \rho K_h \partial \theta_c / \partial t$, giving the vertical heat flux for a vertical wind speed $w$ \citep{Holtslag2009}. We note the resulting vertical surfacic heat flux $Q_h [W.m^{-2}]$. Such models hold the implicit hypothesis of an infinite extension of the city with the same morphology around the studied block so that all horizontal gradients are negligible, thus no horizontal diffusion is considered.
\nomenclature{$\theta_c$}{Canopy potential temperature $[K]$}
\nomenclature{$C_p \rho$}{Thermal capacity of air $[J.K^{-1}.m^{-3}]$}
\nomenclature{$K_h$}{Eddy diffusivity $[m^2.s^{-1}]$}
\nomenclature{$Q_h$}{Vertical turbulent heat flux $[W.m^{-2}]$}

\begin{figure}[tbp]
  \centering
  \includegraphics[clip=true,width=0.99\linewidth]{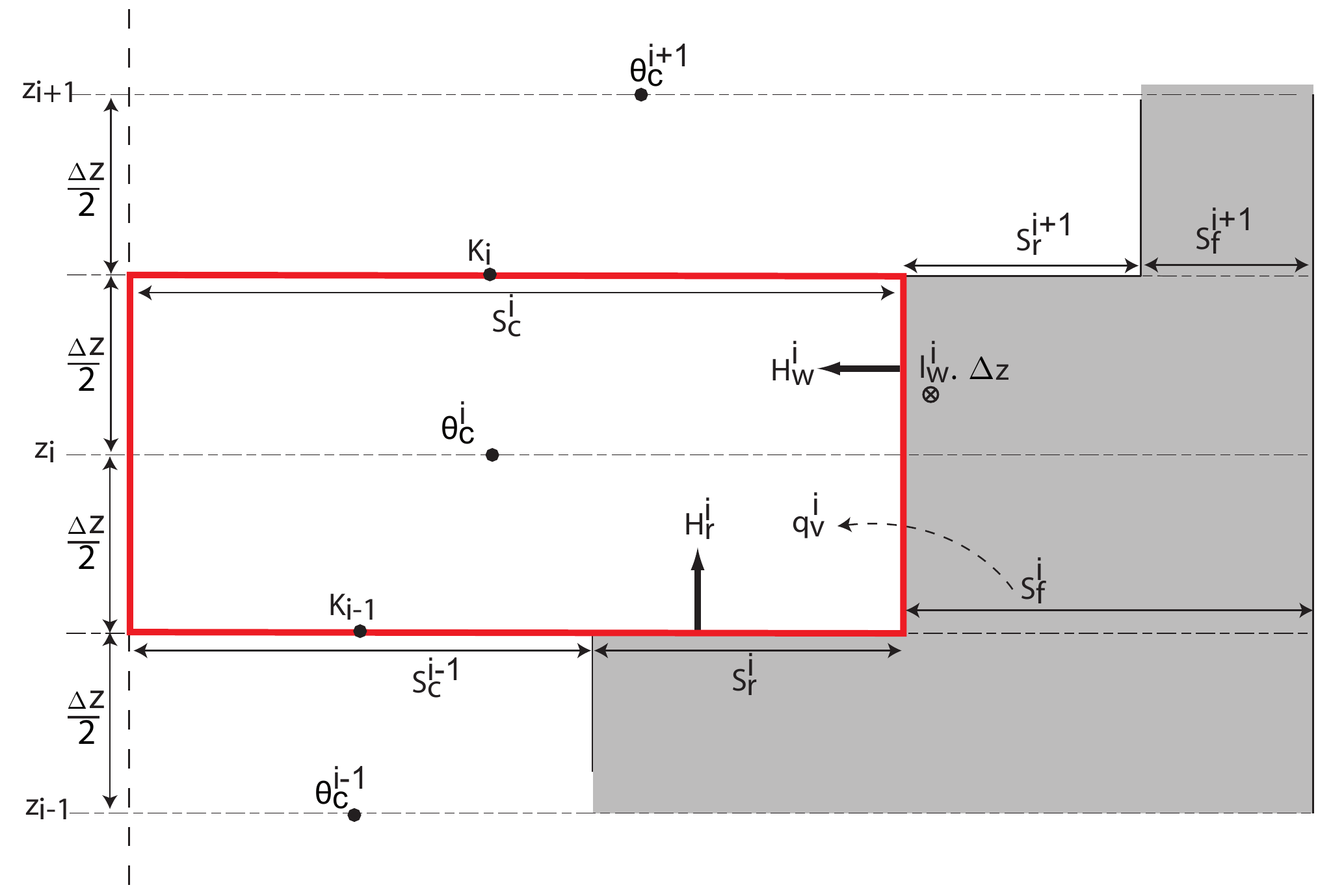}
  \caption{Budget over a layer in the discrete layer scheme. Floor, roof and wall areas, as well as fluxes from them, are cumulative over the horizontal surface of a layer.}
  \label{fig:discretecanopy}
\end{figure}

With this definition of the turbulent flux, we can express the energy budget of a layer of the canopy with an arbitrary shape of buildings. The buildings are described by their cumulative floor area $S_f$ and wall length $l_w$ (sum over all buildings in the layer), giving the free canopy surface in the horizontal plane $S_c$. Roofs are assumed to be only at the intersection of two layers, with surface $S_r$. Inside a layer the characteristics $l_w$ and $S_f$ are assumed to be constant (implying $S_c$ is constant too). We consider four energy fluxes for a layer: surfacic flux from the walls $H_w [W.m^-2]$, surfacic flux from the roofs $H_r[W.m^-2]$, volumic heat flux from the inside air (venting and infiltration) $q_v [W.m^{-3}]$, incoming and outgoing turbulent heat fluxes (Fig. \ref{fig:discretecanopy}). The energy budget inside a layer is then written:
\nomenclature{$S_f$}{Cumulative floor area $[m^2]$}
\nomenclature{$S_r$}{Cumulative roof area $[m^2]$}
\nomenclature{$S_c$}{Free canopy area in the horizontal plan $[m^2]$}
\nomenclature{$l_w$}{Cumulative wall length $[m]$}
\nomenclature{$H_w$}{Heat flux from the walls $[W.m^{-2}]$}
\nomenclature{$H_r$}{Heat flux from the roofs $[W.m^{-2}]$}
\nomenclature{$q_v$}{Heat flux from venting and infiltration $[W.m^{-3}]$}

\begin{align}
\label{eq:canopy}
C_p \rho \frac{\partial \theta_c}{\partial z} &= C_p \rho \frac{\partial}{\partial z} K_h \frac{\partial \theta_c}{\partial z}+H_w \frac{l_w}{S_c} + q_v \frac{S_f}{S_c}
\end{align}

The continuity between two layers at height $z_k$, at the level of roofs, is written:

\begin{align}
\label{eq:continuity}
Q_h(z_k^+)S_{c} &= Q_h(z_k^-)(S_{c}-S_r) + H_r S_{r} \\
\theta_c(z_k^+) &= \theta_c(z_k^-) 
\end{align}

We did not expand the turbulent flux $Q_h$ in eq. \ref{eq:continuity} to avoid defining $K_h$ at the interface of two layers. Note that canopy area $S_c$, floor area $S_f$ and roof area $S_r$ are not independant: between two layers the augmentation of $S_c$ is equal to $S_r$ and is also the reduction of floor area $S_f$ ($\Delta S_c = S_r = -\Delta S_f$).

After establishing the budget inside a layer (eq. \ref{eq:canopy},) we further simplified it with a spatial discretization, with one node per layer (i.e. one canopy temperature per layer, no continuous variation). This discretization facilitates coupling with the BEM (sec. \ref{sec:BEM}), which uses only one outside temperature per layer. It also further simplifies the analytical formulation from one partial differential equation and two continuity conditions to one ordinary differential equation per layer.

The discretization point is the center of the considered layer. We further assume that heat fluxes $H_w$ and $q_v$ depend only on the building from which they originate, and are thus independant of $z$ over the height of a layer (but vary from one layer to another). Integration of eq. \ref{eq:canopy} over a layer of height $\Delta z$, delimited by heights $z_i$ and $z_{i+1}$, gives:
\nomenclature{$\Delta z$}{Height of a layer $[m]$}

\begin{align}
\label{eq:intergation}
C_p \rho \frac{\partial \theta_c^i}{\partial z} &= \frac{(Q_h(z_{i}^+) - Q_h(z_{i+1}^-))}{\Delta z} + H_w \frac{l_w}{S_c} + q_v \frac{S_f}{S_c}
\end{align}

We discretize the turbulent flux using an upwind explicit scheme :  
\begin{align}
\label{eq:upwind}
Q_h(z^{i-}) &= C_p \rho \frac{K_{h}^{i+1} \theta_c^{i+1} + K_{h}^{i-1} \theta_c^{i-1} - 2 K_{h}^i \theta_c^i}{\Delta z}
\end{align}
With $K_{h}^i$ and $\theta_c^i$ respectively the diffusivity and temperature at the center of the layer (also being the average over the height of layer $i$).

Substituting $Q_h(z_i^+)$ using eq. \ref{eq:continuity} and then expanding the turbulent flux using eq. \ref{eq:upwind}, eq. \ref{eq:intergation} then gives:
\begin{align}
\label{eq:discrete}
C_p \rho \frac{d \theta_c^i}{dt} =&  C_p \rho K_h^i \frac{\theta_c^{i+1}-\theta_c^i}{\Delta z^2} \left( 1 - \frac{S_r}{S_c} \right)  + C_p \rho K_h^{i-1} \frac{\theta_c^{i}-\theta_c^{i-1}}{\Delta z^2} \\ \nonumber
& + \frac{H_r S_r }{S_c \Delta z} + H_w \frac{l_w}{S_c} + q_v \frac{S_f}{S_c}
\end{align}

This scheme is pictured in Fig. \ref{fig:discretecanopy}. This discrete layer formulation makes it possible to have only one ordinary differential equation per layer. With the hypothesis of contsant morphological characteristics inside a layer, we were also able to simplify the porosity in the diffusion term. The heat fluxes of the buildings, $H_w$, $H_r$ and $q_v$ depend on the indoor air temperatures and are determined with the BEM (eq. \ref{eq:bem-air}, \ref{eq:bem-mass}). The turbulent diffusivity $K_h$ is then the last unknown.

\subsubsection{Determination of eddy diffusivity}
Once we had established the constitutive equation for the canopy (eq. \ref{eq:discrete}), we needed to compute the eddy diffusivity $K_h(z)$ to solve this energy budget. We chose the linear-exponential approximation \citep{Jerevic2009} for its ease of use and low computation requirements. We considered that a classical use of a planetary boundary layer scheme (\citep{Bougeault1989} in \citep{Martilli2002} or \citep{Mellor1974} in \citep{Kondo2005}) was too computationally intensive for the design of a few urban blocks, with no intention of studying the full mesoscale feedback. The linear-exponential form is written: 
\begin{align}
K_h = (K_{max}e^{1/2}/h_{max})z exp(-0.5(z/h_{max})^2)
\end{align}
Input parameters $K_{max}$ and $h_{max}$ were determined with simulations or experimental databases (in this application we used values from \citep{Jerevic2009}).
\nomenclature{$K_{max}$}{Maximum eddy diffusivity $[m^2.s^{-1}]$}
\nomenclature{$h$}{Maximum eddy diffusivity $[m]$}

Such an approach holds the implicit assumption that $K_h(z)$ is independant from design choices. While urban morphology and eddy diffusivity are linked from a physical point of view, to the best of our knowledge there is no way to explicitely link the two. This conventional approach is in our opinion the best suited to design uses as it allows easier comparisons with less moving parameters, although it is less precise than classical meteorological studies. 
\subsection{Building Energy Model (BEM)}
\label{sec:BEM}
In this section we develop a single zone BEM in order to compute the anthropogenic heat fluxes ($H_w$, $H_r$ and $q_v$) as well as the heating and cooling demands of buildings. It is important that the model require minimal input data to ensure that it is usable in early stage design without making any hypotheses on further stages (architecture and system design mainly) as well as to reduce computation time at the urban scale. We chose a two-node electrical equivalent model, similar to \citep{Nielsen2005} as a starting base. Such models have been developed for early stage design purposes. They have been successfully used at the urban scale coupled to a full radiosity model \citep{Kampf2007} and can be easily coupled to a canopy model \citep{Ihara2008a,Salamanca2009a,Bueno2012b}. Two specific goals must be matched for our specific case: 
\begin{itemize}
\item distinguishing wall and roof heat fluxes since they are realeased in different canopy layers,
\item using no internal or external wall surfaces for the calculation of radiative flux, since they are not defined at such an early stage (the envelope is not yet real surfaces, just a virtual boundary).
\end{itemize}

\begin{figure}[tbp]
  \centering
  \includegraphics[clip=true,width=0.99\linewidth]{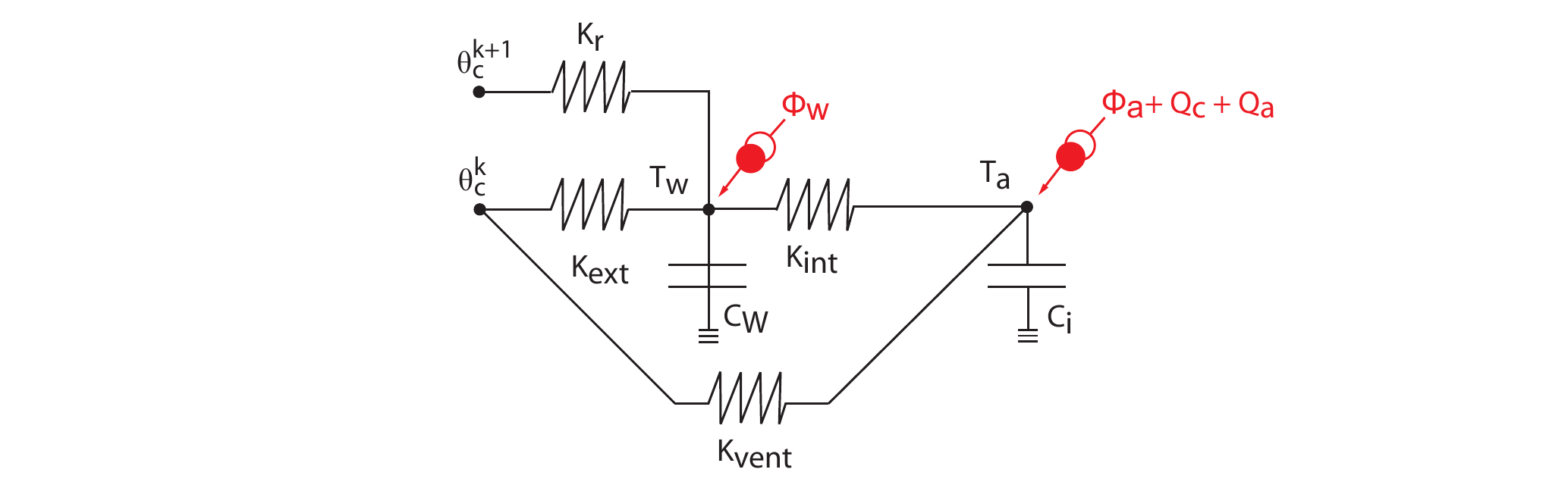}
  \caption{Electrical equivalent of the BEM.}
  \label{fig:batelec}
\end{figure}

Two-node models consider only one thermal mass node for heavy elements (mainly walls and floors) and one inside node (air, furnitures etc.). On the thermal mass node, of temperature $T_w$ and capacity $C_w$, we consider: conduction and convection to the outdoor air through the walls with conductance $K_{ext}$, conduction and convection to the outdoor air through the roofs with conductance $K_{r}$, conduction and convection to the inside air with conductance $K_{int}$ and equivalent radiative flux on the thermal mass $\Phi_w$. Surface convection is embedded in global coefficients $K_{ext}$, $K_{r}$ and $K_{int}$ and is not explicitly computed. On the inside air node, of temperature $T_a$ and capacity $C_a$, we consider: conduction and convection to the thermal mass node with conductance $K_{int}$, direct exchange with the outdoor air (venting, infiltration, windows conduction) with conductance $K_{vent}$, climatic heat flux (heating or cooling, algebraic) $Q_c$, indoor anthropogenic releases $Q_a$ and equivalent radiative flux heating the inside air $\Phi_a$. The full budget for the two nodes, for a zone in the $k-th$ canopy layer, is then written (Fig. \ref{fig:batelec}):
\nomenclature{$T_w$}{Thermal mass temperature $[K]$}
\nomenclature{$T_a$}{Indoor air temperature $[K]$}
\nomenclature{$C_w$}{Thermal mass heat capacity $[J.K^{-1}]$}
\nomenclature{$C_w$}{Indoor air heat capacity $[J.K^{-1}]$}
\nomenclature{$K_{ext}$}{Conductance from thermal mass to outdoor air through walls$[W.K^{-1}]$}
\nomenclature{$K_r$}{Conductance from thermal mass to outdoor air through roof $[W.K^{-1}]$}
\nomenclature{$K_{vent}$}{Conductance from thermal mass to outdoor air throught venting, inflitration and window conduction $[W.K^{-1}]$}
\nomenclature{$Q_a$}{Internal loads $[W.m^{-2}]$}

\begin{align}
\label{eq:bem-mass}
K_{int}(T_a - T_w) + K_{ext}(\theta_c^k - T_w) + K_r(\theta_c^{k+1}-T_w) + \Phi_w &= C_w \frac{d T_w}{dt} \\
\label{eq:bem-air}
K_{int}(T_w - T_a) + K_{vent}(\theta_c^k - T_a) + Q_c + Q_a + \Phi_a & = C_a \frac{d T_a}{dt}
\end{align}

Those equations include canopy temperatures of layers $k$ and $k+1$, $\theta^k_c$ and $\theta^{k+1}_c$. It is thus required to solve eq. \ref{eq:discrete} at the same time as the above BEM's equations, as depicted in the workflow (sec. \ref{sec:workflow}. 

To compute the equivalent radiative flux on the internal air $\Phi_a$, we define the surfacic opening coefficient $\eta$ such as $\Phi_a = \eta \Phi$, where $\Phi$ is the total net radiative flux. $\eta$ includes the areas of both openings, their material charcteristics and various geometric effects (window frames, corners, etc.) at once, with no need to further specify such details. The equivalent radiative flux on the thermal mass is defined by $\Phi_w = \Phi (1 - \eta) K_{ext}/h_{conv}$ \footnote{This formula is easily deduced from the electric equivalent, considering a separate resistance for the convection.}, with $h_{conv}$ the convection coefficient.
\nomenclature{$\Phi$}{Net radiative flux on the external surface $[W.m^{-2}]$}
\nomenclature{$\Phi_w$}{Net radiative flux on the thermal mass $[W.m^{-2}]$}
\nomenclature{$\Phi_a$}{Net radiative flux on the indoor air $[W.m^{-2}]$}
\nomenclature{$\eta$}{Opening coefficient}

We use a simple constant-linear control scheme for the climatic flux $Q_c$, because it is very simple. We make no assumption on the energetic system used but allow for inside air temperature fluctuation. The bounding powers $Q_c^{max}$ (maximum heating) and $Q_c^{min}$ (maximum cooling) are assumed to happen at $\pm 2K$ around targ temperature $T_{targ}$, viz. : $Q_c = Q_c^{min} + ( Q_c^{max} - Q_c^{min} ) (T_a - T_{targ} - 2)/4 $. 
\nomenclature{$Q_c$}{Climatic heat flux $[W.m^{-2}]$}
\nomenclature{$Q_c^{min}$}{Climatic heat flux lower bound (algebraic) $[W.m^{-2}]$}
\nomenclature{$Q_c^{max}$}{Climatic heat flux upper bound (algebraic) $[W.m^{-2}]$}
\nomenclature{$T_{targ}$}{Target indoor temperature $[W.m^{-2}]$}

All the equivalent characteristics ($K_{ext}$, $K_{int}$, $K_r$, $K_{vent}$, $\Phi_w$, $\Phi_a$) allow us to work directly on urban scale parameters, without making hypotheses on smaller scales (e.g. wall composition, internal zoning).
\subsection{Radiosity method}
After establishing the energy budget for the canopy model (eq. \ref{eq:canopy}) and the BEM (eqs. \ref{eq:bem-mass} and \ref{eq:bem-air}), we use a radiosity method to determine the net radiative flux $\Phi$ -- the last unknown of the system. Radiosity methods make it possible to compute both longwaves and shortwaves ranges and to account for all reflexions, on all possible geometries. The output is then the total net radiative flux for all wave lengths, including effect of all obstructions and reflection on ground

We applied the radiosity method with the CitySim software (as opposed to an analytical formulation for the canopy model and BEM) because there was no practical interpretation of the analytical formula for design at the current time and we had no need to improve the analytical form. Moreover, radiosity methods require computing forms factors, a computationally intensive task which is sped up by the use of the simplified radiosity algorithm \citep{Robinson2005} in CitySim.
\subsection{Inter-models workflow}
\label{sec:workflow}
\begin{figure}[tbp]
  \centering
  \includegraphics[clip=true,width=0.99\linewidth]{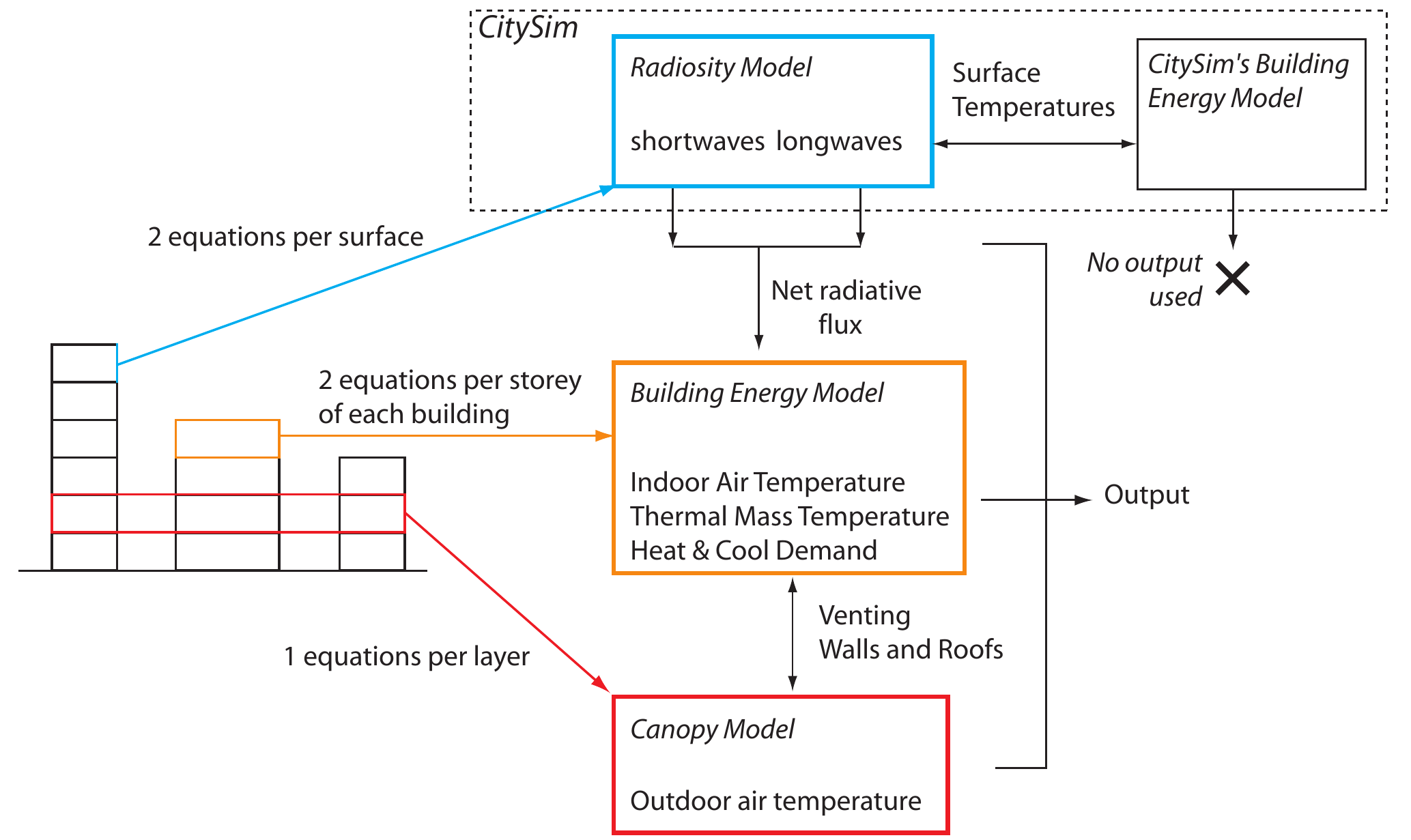}
  \caption{Inter-model workflow.}
  \label{fig:workflow}
\end{figure}

Once all the submodels had been established, we assembled them into an integrated model so that  all the feedbacks were modeled. We used one independant BEM for each storey of each building (as opposed to one equivalent building for all the scenes in most micro-climate studies). This enabled us to observe the effect of a full radiative budget on the real geometry, not only on heat demand, but also on the micro-climate feedback. The data are exchanged between the analytical models (canopy, BEM) and CitySim with an Xquerry script, since CitySim uses a hierchical XML description of the morphology very similar to the one we use for our canopy model and BEM. The final system has then two equations for each surface for the radiative scheme (longwaves and shortwaves), two equations for each storey of each building for the BEM and one equation for each layer of the canopy (Fig. \ref{fig:workflow}).

\section{Test : regular and irregular morphologies}
With the previously established model (sec. \ref{sec:eqs}), we tested four theoritical morphologies, in order to investigate the effect of 3D morphological varations on the feedback loops between outdoor air, indoor air and radiative budget.
\subsection{Case descriptions}
\begin{figure}[tbp]
  \centering
  \includegraphics[clip=true,width=0.99\linewidth]{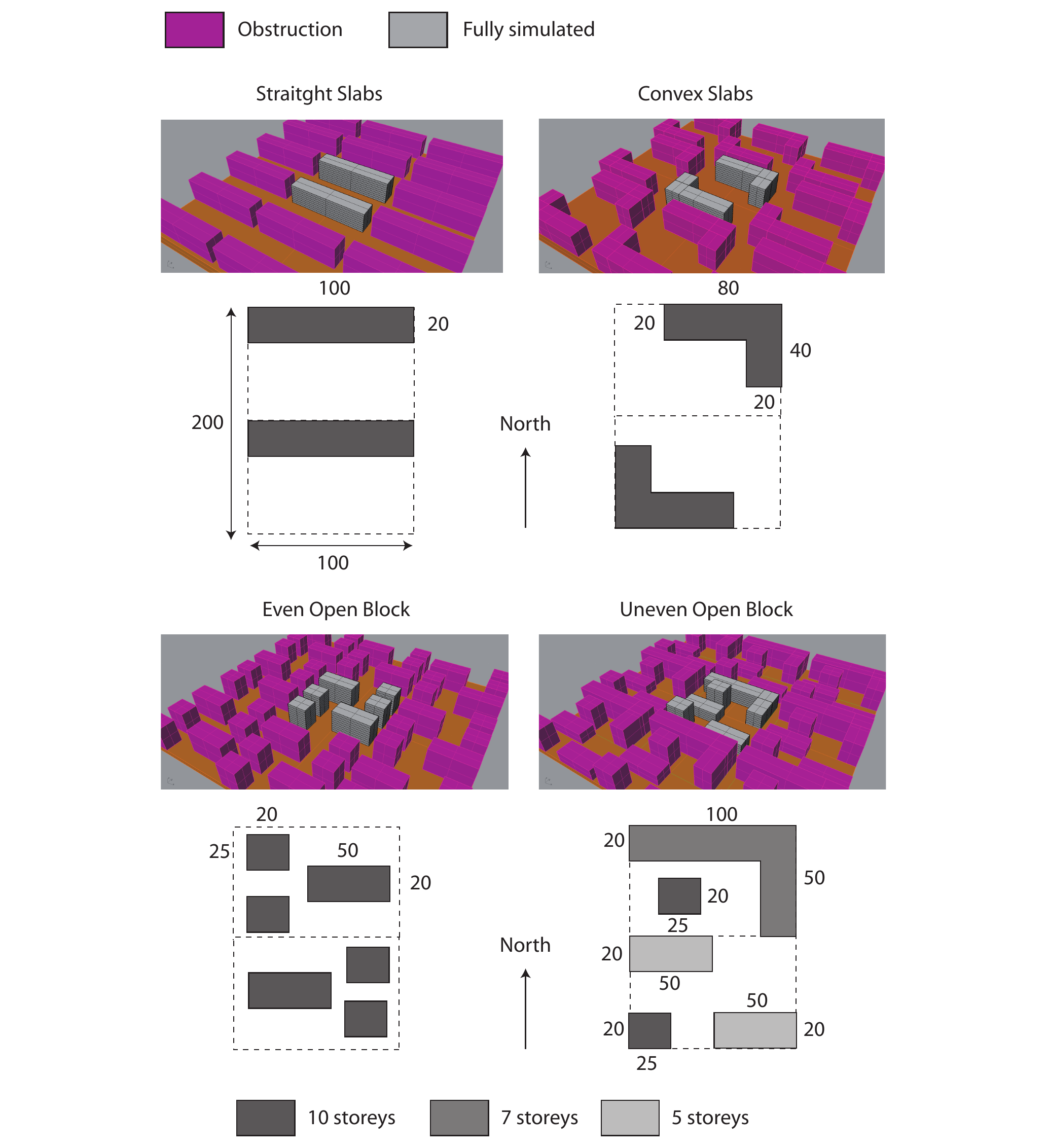}
  \caption{Studied morphologies description.}
  \label{fig:morpho}
\end{figure}

The four tested morphologies were: regular slabs, convex slabs, even open block (all buildings have the same height), uneven open block (Fig. \ref{fig:morpho}). We used the regular slabs as a basis for relative comparisons which made it possible to study three morphological transformations: bending, division into several buildings and modification of the vertical density gradient (roof lines distribution).

All cases are of equal density, with $40.000 m^2$ of floor area, on a $100*200 m^2$ plot. The tallest buildings are always 10 storeys of $3m$, placing the canopy line at $30m$. Moreover, we maintain a width (smallest horizontal dimension) of $20 m$ for all buildings in all cases, to make sure that the contribution of solar gains to the indoor temperature is affected only by the above mentioned transformations. Such width can be considered average, leading to a low, but not uncommon, passive volume ratio \citep{Ratti2005}. It should \emph{a priori} ensure that the effect on solar access and the direct effect of the outdoor temperature are both significant in the observed results.

We use thermal charactestics of standard buildings of the 2000'ʂ, taken from \citep{Kampf2007} under the Basel climate (also used in \citep{Robinson2006}). For the sake of simplicity, we use a constant internal loads throughout the day. Eddy diffusivity is taken from \citep{Jerevic2009} (values summarized in Tab. \ref{tab:therm}). We use values from the literature in order to make comparisons as easy as possible and to focus on the effect of the morphology. We study a winter day (3rd of January), since it should demonstrate high sensitivity to differences between solar gains and heat demand. Multilayer canopy models can also simulate the winter temperature inversion \citep{Kanda2005}, a much less studied phenomena than its summer counterpart.

We ran simulations on a 3x3 grid, with the central cell containing the studied buildings and the surrounding cells being replica of the studied morphology in order to create the obstructions from the surrounding city. The turbulent diffusion is simulated up to $60m$ -- i.e double the canopy line -- to compute the perturbation induced by the urban area above the canopy line, while keeping a reasonable height/width for the simulated column. We force the temperature to the undisturbed atmospheric temperature at $60m$ as a boundary condition.

\begin{table}
	\centering
	\begin{tabular}{llll}
	\toprule
	Conductance per unit of area & $K_{ext}$ & $K_{int}$ & $K_{vent}$  \\
	$[W.K^{-1}.m^{-2}]$ & 0.15 & 0.15 & 0.3  \\  
	\midrule
	Heat capacity per unit of area & $C_{w}$ & $C_{int}$ & \\
	$[J.K^{-1}.m^{-2}]$ & 3E5  & 1.4E3 &  \\
	\midrule
	Powers & $Q_{min}$ & $Q_{max}$ & $Q_{a}$ \\
	$[W.m^ {-2}]$ & -100 & 100 & 100 \\
	\midrule
	 Albedo & $\alpha$\\
	 & 0.2  \\
	\midrule
	 Eddy Diffusivity & $K_{max}$ & $h$ &  \\
	 & 2 $[m^2.s^{-1}]$ & 60 $[m]$& \\	
	\bottomrule
	\end{tabular}
	\label{tab:therm}
	\caption{Thermal characteristics for the test cases.}
\end{table}

\subsection{Results}
We first studied the solar gains and surface dissipation (losses through walls and roofs) as both should increase with the surface envelope, leading to a compromise for heat demand. After those three components, we studied the canopy temperature distribution, both for its effect on heat demand and for the urban heat island effect.

\begin{figure}[tbp]
  \centering
  \includegraphics[clip=true,width=0.99\linewidth]{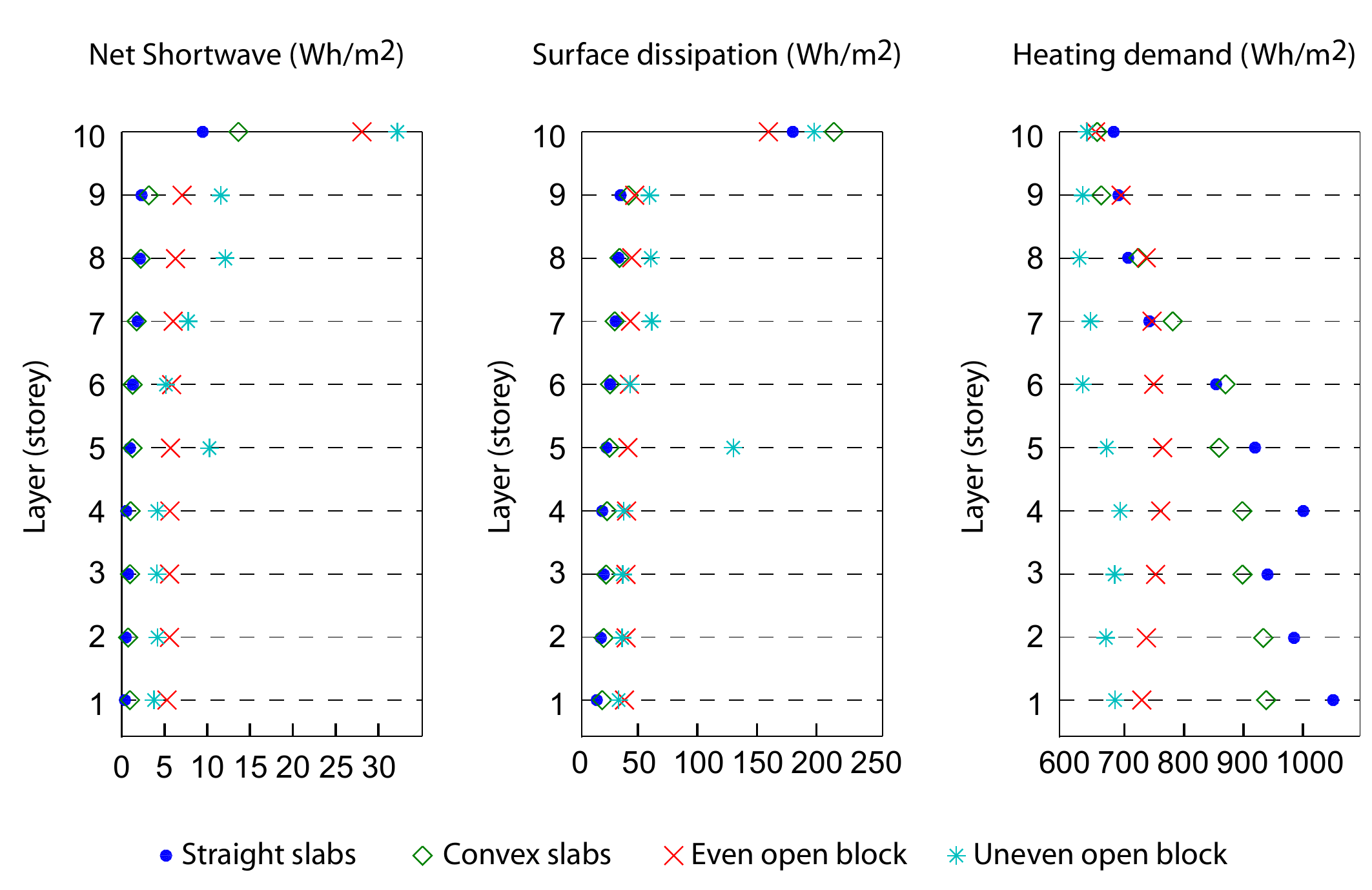}
  \caption{Cumulative sums on all buildings of a layer and on 24 hours of solar irradiance, surface dissipation and heat demand, normalized by floor area in the layer.}
  \label{fig:energy}
\end{figure}

We present the results in the form of total energy over the 24 hours, summed over all the building of a layer and then normalized by the floor area of the layer. This apporach makes it possible for us to focus specifically on the vertical design with a single value per layer, while still acounting for uneven floor area distribution (Fig. \ref{fig:energy}).

\subsubsection{Solar access}
For the solar gains, we choose to display only the shortwave net flux (including reflexions) on Fig. \ref{fig:energy}, as it best reflects sun and sky access, while the longwaves range also partly reflects surface dissipation and heat storage (longwaves are included in the calculation of heat demand nonetheless). On an average, solar radiations are 7 times higher on even and uneven open blocks than on regular slabs. Such a high factor is  due mainly to the very low values for the regular slabs. Even and uneven open blocks display very similar results on average, with only a $4\%$ difference, but can display up to $100\%$ difference on specific layers due to roof surface variations. Convex slabs display a $40\%$ increase over regular slabs due to improved sun access with the large opening of the pseudo-courtyard and to the reflexions on the convex angles. In terms of vertical distribution, the relative increase for even and uneven open blocks compared to both slab cases is higher in the lower levels. Overall, we can clearly distinguish two groups: morphologies with large envelope surfaces (open blocks) and high solar gains  -- especially in the lower levels -- and morphologies with small envelope surfaces (slabs) and low solar gains.

\subsubsection{Surface dissipation}
We define the surface dissipation of a layer as the sum of $H_w$ -- released in the current layer -- and $H_r$ -- released in the next layer -- from the buildings of the current layer. It represents energy losses from surfacic exchanges for buildings of a given story. It must not be confused with the anthropogenic heat released in the layer -- $H_w$ from the current layer and $H_r$ from the previous layer, both released in the current layer. The surface dissipation is $11\%$ higher for the convex slabs than for the straight slabs, despite the same envelope surface for both cases. This difference may be explained by a higher thermal mass temperature due to increased solar gains for the convex slabs, leading to a higher temperature gradient, but also by microclimatic effects. As expected, the even and uneven open blocks respectively display a $25\%$ and $56\%$ increase compared to the regular slabs because they have much larger dissipation surfaces. Layers with roofs display much larger dissipation, with a factor exceeding $400\%$. The effect is almost constant on the vertical axis for morphologies with even floor area distributions -- except for the last layer due to the roof. However, the surface dissipation (per square meter of floor area in the considered layer) can vary by a factor of up to 4.5 between layers of the uneven block. As opposed to the other cases, the uneven block also displays variations of up to $60\%$ between layers with no roofs (i.e. excluding layers 5,7 and 10). Overall, we can distinguish the same two groups for the surface losses as for the solar gains, although the effects are far from being perfectly symmetric.

\subsubsection{Heat demand}
We define heating and cooling demands as $Q_c$ in absolute values ($Q_c$ is negative for cooling). We expect this heat demand to be a balance between the increased solar gains and increased losses, for both even and uneven open blocks. Results show that overall the solar gains weigh more in the balance. Convex slabs, even open blocks and uneven open blocks, show respectively $5\%$, $15\%$ and $25\%$ decreases in heating and cooling demands compared to straight slabs. The lower we go in the layers, the stronger the effect. The effect is not perfectly linear, with only a $7\%$ difference between straight slabs and uneven open blocks in the 10th layer, but a $35\%$ difference in the first. The open block with an uneven floor space distribution performs better in every single layer and achieves almost constant performance in every layer, with a maximum variation of only $10 \%$ between layers, as opposed to $35 \%$ for the straight slabs.

\subsubsection{Microclimate}
\begin{figure}[tbp]
  \centering
  \includegraphics[clip=true,width=0.99\linewidth]{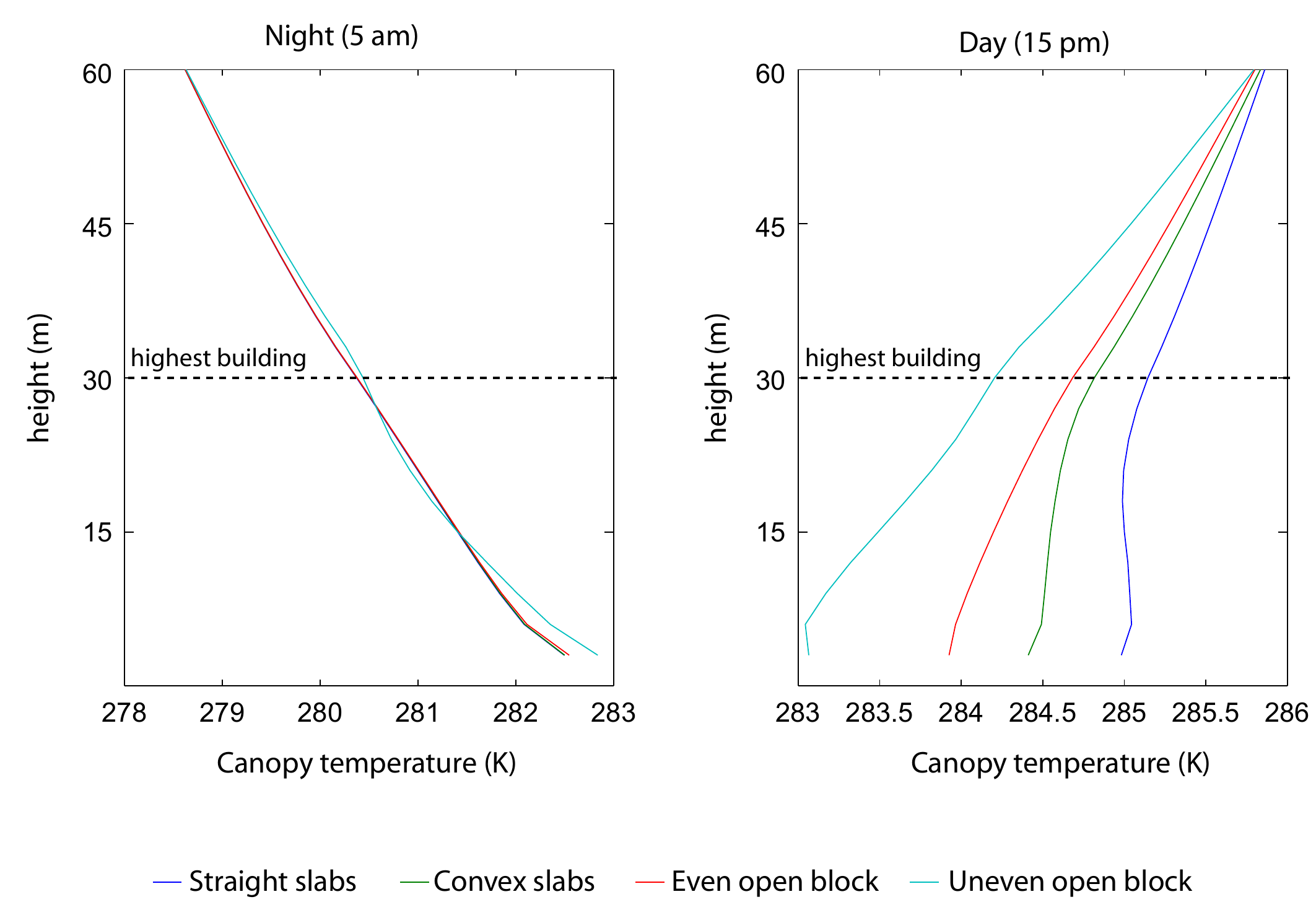}
  \caption{Vertical temperature profile, during night (5am) and day. First 30m are the urban canopy (under roof line), 30 to 60m is simulated above the canopy.}
  \label{fig:climate}
\end{figure}

The effect of morphology on the canopy temperature during the night is almost null. At 5am (Fig. \ref{fig:climate}), we observe a regular gradient, almost identical for all morphologies, with a mean deviation between cases of less than $0.1 \%$. The temperature is higher in the bottom layers and lower in the top layers, with a difference of $ 1.7K$ -- respectively $3.6K$ -- between the first layer -- respectively the 10th layer (end of the urban canopy)-- and the atmospheric temperature (undisturbed temperature). This temperature gradient corresponds to a constant upward turbulent flux, thus constant anthropogenic heat releases.

During the day we observe a reverted temperature gradient for all cases, with the highest temperatures in the top layer and the lowest in the bottom layer. As opposed to the night, the effect of morphology on the microclimate is significant at 3pm. The uneven open block displays an almost constant temperature gradient, i.e a constant turbulent flux, creating a high heat circulation. The regular slabs display an almost constant temperature underneath the urban canopy line, meaning heat is stacking in the lower layers. The different dynamics result in significant differences in all layers, with a maximum difference of $1K$ between cases in the 10th layer (top of the urban canopy) and a maximum difference of $2K$ between cases in the first layer. We observe a strict hierarchy between cases, with straight slabs always having higher canopy temperatures than convex slabs. Then come even open blocks and uneven open blocks. This hierarchy is exactly the same as the one for heat demand, i.e. the morphologies with the highest canopy temperatures are those with the highest energy input -- an expected result.
\section{Discussion on the model and the results of the test cases}
We originally assumed that existing physics was sufficient to study energy flow in 3D urban fabrics, but needed to be restated around a common description of the urban morphology in order to be usable for design purposes. We were able to produce such a model, with new analytical developments and hypotheses. Since the core physic of the model is roughly the same as the one of the initial models, the results should present the same degree of confidence. However, the problem of numerical values in early stage design remains, as they are chosen on very rough estimate. The presented cases should then only be considered as theoretical.

The first of these hypotheses was to consider that the urban geometry is composed of layers of constant characteristics (both thermal and morphological) and only vertical and horizontal walls, i.e. the morphology is described story by story with \emph{shoe box} elements. First, this hypothesis was taken from urban master plans, where shoe box representations are widely used. Second, given the vast array of possible morphologies to be described in this way, we are presently convinced that this hypothesis allows for greater generality and is suitable for most design applications. 

The second hypothesis we had to make was the climatic forcing, in the form of a forced atmospheric temperature and a forced eddy diffusivity profile. Temperature forcing at the top of the studied domain is common practice in urban climate studies -- commonly referred to as off-line simulation \citep{Lemonsu2004,Hamdi2008} -- especially for relative comparisons. However, by forcing a constant diffusivity profile we neglected all the time variations of the diffusivity (especially day-night variations) and the effects of morphology on diffusivity. The time variation could be compensated with better data, or even by coupling with a mesoscale model, although the latter method is too heavy for simple engineering needs in our opinion. To our knowledge, there are currently no methods explicitly linking eddy diffusivity to urban morphology. Given that we make only relative comparisons, and in view of the results, it is likely that this hypothesis does not invalidate the core logic of our results. We are of the opinion that linear-exponential profiles are a good simplification, especially for engineering and design applications, but will require more work to be fully usable on real cases.

\citep{Ratti2005} suggested that urban morphology can change energy demands by a factor of about two. We were indeed able to confirm  this order of magnitude of two on heat demand. Studies at constant density and thermal characteristics, but varying morphologies, remain very rare, although studies like the present one prove they can be very insightful. One reason is that they require specific models. Our model can also simulate the winter temperature inversion as observed by \citep{Kanda2005}. On the basis of our results, we are presently convinced that more studies on the winter urban micro-climate could lead to new innovative solutions. Once again, such a task requires a 3D model to simulate the vertical temperature gradient and the effect of obstructions. Indeed, most of our cases affect only obstruction parameters, but the net radiative fluxes then have significant effects on heat demand and the urban climate through feedback loops, stressing the need for fully coupled models. 

Our last and main initial hypothesis, was that developing a design model would dramatically change its use and interpretation compared to the starting models -- even if the core physics are almost the same. Due to the high number of feedback loops and the consequent non-linearity, one can hardly anticipate the effect of a specific parameter change. The core logic of our model is then case-based reasoning \citep{Gilboa2014}. The user builds a specific case, simulates it and uses it as an analogy that helps him reason, hints at design solutions or examples contrary to general rules or beliefs, but does not try to generalize the results into a hard rule. This design parameter approach can be considered the opposite of the morphological indicator where one tries to infer a general rule from a parameter supposed a priori, not necessarily related to a specific analytical formulation. It is our strong belief that the morphological indicator approach (e.g. \citep{Steemers2003,Robinson2006}), while good for evaluating existing urban fabrics, is not suited to design purposes. The indicator approach would force the designer into a reduced spectrum of solutions (e.g. a range of surface to volume ratio), excluding a priori some particular solutions. Our results show that some original solutions are worth considering and can be very challenging to both the intuition and hard rules: the case with the highest surface to volume ratio has the lowest heat demand but produces the strongest micro-climate distortion\footnote{In the form of temperature inversion, not temperature increase}. In the parameter approach, virtually every imaginable solution is potentially worth testing, especially the most original ones. The role of a model like the one presented in this study is then to test and back up the logic of the designer on new solutions. The model itself can even suggest ways of producing new solutions by testing new values for parameters. This change from a rule-based to a case-based use and interpretation represents a switch from an assessment goal to a design goal.
\section{Conclusion}
In the course of the present study we produced a new urban design energy model and an application on four urban morphologies. First, starting with existing models of urban physics, we were able to produce a design model of coupled urban canopy, BEM and radiosity for three dimensional urban forms in order to study coupled indoor and outdoor energy transfers during the urban design process. We built the parameters of the equations to be simple design parameters (floor or wall areas, global conductance, etc.) shared by all three sub-models and revolving around a description of the morphology as close as possible to urban design practice. One could say urban design and physics speak the same language in our model. As an example use of this model, we showed that three dimensional urban morphology design choices, such as convexity, number of buildings or height gradient, could reduce heat demand by a factor of two and make heat demand almost even between all levels, while a bad design choice lead to almost double the heat the demand for lower levels than for top levels. Those morphology choices can also amplify the winter temperature inversion effect, with a magnitude of about $2K$. It is noteworthy that the case with the best sun access and the lowest heat demand also lead to higher climate distortion due to effect of the turbulent diffusion. In light of these results on very simple variations, we find every reason to believe that the study of more complex morphologies could unveil unknown effects and lead to creative design solutions to energy problems. It is our firm opinion that more research is needed on integrated tools that allow architects, engineers and physicists to think and communicate with a common language and shared objects so as to reach real interdisciplinary action on the critical problem of energy in urban planning. 
\section*{Acknowledgements}
This research is part of the SERVEAU project, funded under the Single-Inter-Ministry fund.
We sincerely thank Dr. J.Kämpf and the LESO-PB laboratory for the training on the CitySim software. We sincerely thank the reviewers for their voluntary help in making the document more clear.

\bibliographystyle{elsarticle-num}
\bibliography{EB2014}

\begin{thebibliography}{10}
\expandafter\ifx\csname url\endcsname\relax
  \def\url#1{\texttt{#1}}\fi
\expandafter\ifx\csname urlprefix\endcsname\relax\def\urlprefix{URL }\fi
\expandafter\ifx\csname href\endcsname\relax
  \def\href#1#2{#2} \def\path#1{#1}\fi

\bibitem{Firley2009}
E.~Firley, C.~Stahl, {The urban housing handbook}.

\bibitem{Oke1982}
T.~Oke, {The energetic basis of the urban heat island}, Quarterly journal of
  the royal meteorology society 108~(455).

\bibitem{Masson2000}
R.~Y. Masson, {A physically-based scheme for the urban energy budget in
  atmospheric models}, Boundary-Layer Meteorology~(September 1999) (2000)
  357--397.

\bibitem{Rasheed2011}
A.~Rasheed, D.~Robinson,
  \href{http://onlinelibrary.wiley.com/doi/10.1002/joc.2240/full}{{Representing
  complex urban geometries in mesoscale modeling}}, International Journal
  \ldots.
\newline\urlprefix\url{http://onlinelibrary.wiley.com/doi/10.1002/joc.2240/full}

\bibitem{Martilli2002}
A.~Martilli, A.~Clappier, M.~Rotach,
  \href{http://www.springerlink.com/index/FDEUV1UY3N9W7AK4.pdf}{{An urban
  surface exchange parameterisation for mesoscale models}}, Boundary-Layer
  Meteorology 104~(2) (2002) 261--304.
\newline\urlprefix\url{http://www.springerlink.com/index/FDEUV1UY3N9W7AK4.pdf}

\bibitem{Hamdi2008}
R.~Hamdi, V.~Masson,
  \href{http://journals.ametsoc.org/doi/full/10.1175/2008JAMC1865.1}{{Inclusion
  of a Drag Approach in the Town Energy Balance (TEB) Scheme: Offline 1D
  Evaluation in a Street Canyon}}, Journal of Applied Meteorology and
  Climatology 47~(10) (2008) 2627--2644.
\newblock \href {http://dx.doi.org/10.1175/2008JAMC1865.1}
  {\path{doi:10.1175/2008JAMC1865.1}}.
\newline\urlprefix\url{http://journals.ametsoc.org/doi/full/10.1175/2008JAMC1865.1}

\bibitem{Bueno2012b}
B.~Bueno, L.~Norford, G.~Pigeon, R.~Britter,
  \href{http://dx.doi.org/10.1016/j.buildenv.2012.01.023}{{A
  resistance-capacitance network model for the analysis of the interactions
  between the energy performance of buildings and the urban climate}}, Building
  and Environment 54 (2012) 116--125.
\newblock \href {http://dx.doi.org/10.1016/j.buildenv.2012.01.023}
  {\path{doi:10.1016/j.buildenv.2012.01.023}}.
\newline\urlprefix\url{http://dx.doi.org/10.1016/j.buildenv.2012.01.023}

\bibitem{Salamanca2009a}
F.~Salamanca, A.~Martilli,
  \href{http://www.springerlink.com/content/n87065mu46138773/}{{A new Building
  Energy Model coupled with an Urban Canopy Parameterization for urban climate
  simulations—part II. Validation with one dimension off-line simulations}},
  Theoretical and Applied Climatology 99~(3-4) (2009) 345--356.
\newblock \href {http://dx.doi.org/10.1007/s00704-009-0143-8}
  {\path{doi:10.1007/s00704-009-0143-8}}.
\newline\urlprefix\url{http://www.springerlink.com/content/n87065mu46138773/}

\bibitem{Ihara2008a}
I.~Tomohiko, Y.~Kikegawa, K.~Asahi, Y.~Genchi, H.~Kondo,
  \href{http://linkinghub.elsevier.com/retrieve/pii/S0306261907000888}{{Changes
  in year-round air temperature and annual energy consumption in office
  building areas by urban heat-island countermeasures and energy-saving
  measures}}, Applied Energy 85~(1) (2008) 12--25.
\newblock \href {http://dx.doi.org/10.1016/j.apenergy.2007.06.012}
  {\path{doi:10.1016/j.apenergy.2007.06.012}}.
\newline\urlprefix\url{http://linkinghub.elsevier.com/retrieve/pii/S0306261907000888}

\bibitem{Grimmond2011}
C.~S.~B. Grimmond, M.~Blackett, M.~J. Best, J.-J. Baik, S.~E. Belcher,
  J.~Beringer, S.~I. Bohnenstengel, I.~Calmet, F.~Chen, A.~Coutts, A.~Dandou,
  K.~Fortuniak, M.~L. Gouvea, R.~Hamdi, M.~Hendry, M.~Kanda, T.~Kawai,
  Y.~Kawamoto, H.~Kondo, E.~S. Krayenhoff, S.-H. Lee, T.~Loridan, A.~Martilli,
  V.~Masson, S.~Miao, K.~Oleson, R.~Ooka, G.~Pigeon, A.~Porson, Y.-H. Ryu,
  F.~Salamanca, G.~Steeneveld, M.~Tombrou, J.~A. Voogt, D.~T. Young, N.~Zhang,
  \href{http://doi.wiley.com/10.1002/joc.2227}{{Initial results from Phase 2 of
  the international urban energy balance model comparison}}, International
  Journal of Climatology 31~(2) (2011) 244--272.
\newblock \href {http://dx.doi.org/10.1002/joc.2227}
  {\path{doi:10.1002/joc.2227}}.
\newline\urlprefix\url{http://doi.wiley.com/10.1002/joc.2227}

\bibitem{Robinson2006}
D.~Robinson,
  \href{http://linkinghub.elsevier.com/retrieve/pii/S0038092X06000612}{{Urban
  morphology and indicators of radiation availability}}, Solar Energy 80~(12)
  (2006) 1643--1648.
\newblock \href {http://dx.doi.org/10.1016/j.solener.2006.01.007}
  {\path{doi:10.1016/j.solener.2006.01.007}}.
\newline\urlprefix\url{http://linkinghub.elsevier.com/retrieve/pii/S0038092X06000612}

\bibitem{Ratti2003}
C.~Ratti, D.~Raydan, K.~Steemers,
  \href{http://dx.doi.org/10.1016/S0378-7788(02)00079-8}{{Building form and
  environmental performance: archetypes, analysis and an arid climate}}, Energy
  and Buildings 35~(1) (2003) 49--59.
\newblock \href {http://dx.doi.org/10.1016/S0378-7788(02)00079-8}
  {\path{doi:10.1016/S0378-7788(02)00079-8}}.
\newline\urlprefix\url{http://dx.doi.org/10.1016/S0378-7788(02)00079-8}

\bibitem{Kondo2005}
H.~Kondo, Y.~Genchi, Y.~Kikegawa, Y.~Ohashi, H.~Yoshikado, H.~Komiyama,
  \href{http://www.springerlink.com/content/l0984347216344n4/}{{Development of
  a Multi-Layer Urban Canopy Model for the Analysis of Energy Consumption in a
  Big City: Structure of the Urban Canopy Model and its Basic Performance}},
  Boundary-Layer Meteorology 116~(3) (2005) 395--421.
\newblock \href {http://dx.doi.org/10.1007/s10546-005-0905-5}
  {\path{doi:10.1007/s10546-005-0905-5}}.
\newline\urlprefix\url{http://www.springerlink.com/content/l0984347216344n4/}

\bibitem{Holtslag2009}
A.~A.~M. Holtslag, G.-J. Steeneveld, {Single Column Modeling of Atmospheric
  Boundary Layers and the Complex Interactions with the Land Surface}, in:
  Encyclopedia of Complexity and Systems Science, 2009, pp. 8139--8153.

\bibitem{Jerevic2009}
A.~Jeri\v{c}evi\'{c}, v.~Ve\v{c}enaj,
  \href{http://link.springer.com/10.1007/s10546-009-9367-5}{{Improvement of
  Vertical Diffusion Analytic Schemes Under Stable Atmospheric Conditions}},
  Boundary-Layer Meteorology 131~(2) (2009) 293--307.
\newblock \href {http://dx.doi.org/10.1007/s10546-009-9367-5}
  {\path{doi:10.1007/s10546-009-9367-5}}.
\newline\urlprefix\url{http://link.springer.com/10.1007/s10546-009-9367-5}

\bibitem{Bougeault1989}
P.~Bougeault, P.~Lacarrere,
  \href{http://journals.ametsoc.org/doi/abs/10.1175/1520-0493(1989)117<1872:POOITI>2.0.CO;2}{{Parameterization
  of Orography-Induced Turbulence in a Mesobeta--Scale Model}}, Monthly Weather
  Review 117~(8) (1989) 1872--1890.
\newblock \href
  {http://dx.doi.org/10.1175/1520-0493(1989)117<1872:POOITI>2.0.CO;2}
  {\path{doi:10.1175/1520-0493(1989)117<1872:POOITI>2.0.CO;2}}.
\newline\urlprefix\url{http://journals.ametsoc.org/doi/abs/10.1175/1520-0493(1989)117<1872:POOITI>2.0.CO;2}

\bibitem{Mellor1974}
G.~L. Mellor, T.~Yamada,
  \href{http://journals.ametsoc.org/doi/abs/10.1175/1520-0469(1974)031<1791:AHOTCM>2.0.CO;2}{{A
  Hierarchy of Turbulence Closure Models for Planetary Boundary Layers}},
  Journal of the Atmospheric Sciences 31~(7) (1974) 1791--1806.
\newblock \href
  {http://dx.doi.org/10.1175/1520-0469(1974)031<1791:AHOTCM>2.0.CO;2}
  {\path{doi:10.1175/1520-0469(1974)031<1791:AHOTCM>2.0.CO;2}}.
\newline\urlprefix\url{http://journals.ametsoc.org/doi/abs/10.1175/1520-0469(1974)031<1791:AHOTCM>2.0.CO;2}

\bibitem{Nielsen2005}
T.~R. Nielsen, \href{http://dx.doi.org/10.1016/j.solener.2004.06.016}{{Simple
  tool to evaluate energy demand and indoor environment in the early stages of
  building design}}, Solar Energy 78~(1) (2005) 73--83.
\newblock \href {http://dx.doi.org/10.1016/j.solener.2004.06.016}
  {\path{doi:10.1016/j.solener.2004.06.016}}.
\newline\urlprefix\url{http://dx.doi.org/10.1016/j.solener.2004.06.016}

\bibitem{Kampf2007}
J.~H. K\"{a}mpf, D.~Robinson,
  \href{http://dx.doi.org/10.1016/j.enbuild.2006.09.002}{{A simplified thermal
  model to support analysis of urban resource flows}}, Energy and Buildings
  39~(4) (2007) 445--453.
\newblock \href {http://dx.doi.org/10.1016/j.enbuild.2006.09.002}
  {\path{doi:10.1016/j.enbuild.2006.09.002}}.
\newline\urlprefix\url{http://dx.doi.org/10.1016/j.enbuild.2006.09.002}

\bibitem{Robinson2005}
D.~Robinson, A.~Stone,
  \href{http://bse.sagepub.com/cgi/doi/10.1191/0143624405bt133oa}{{A simplified
  radiosity algorithm for general urban radiation exchange}}, Building Services
  Engineering Research and Technology 26~(4) (2005) 271--284.
\newblock \href {http://dx.doi.org/10.1191/0143624405bt133oa}
  {\path{doi:10.1191/0143624405bt133oa}}.
\newline\urlprefix\url{http://bse.sagepub.com/cgi/doi/10.1191/0143624405bt133oa}

\bibitem{Ratti2005}
C.~Ratti, N.~Baker, K.~Steemers,
  \href{http://linkinghub.elsevier.com/retrieve/pii/S0378778804003391
  http://www.sciencedirect.com/science/article/pii/S0378778804003391}{{Energy
  consumption and urban texture}}, Energy and Buildings 37~(7) (2005) 762--776.
\newblock \href {http://dx.doi.org/10.1016/j.enbuild.2004.10.010}
  {\path{doi:10.1016/j.enbuild.2004.10.010}}.
\newline\urlprefix\url{http://linkinghub.elsevier.com/retrieve/pii/S0378778804003391
  http://www.sciencedirect.com/science/article/pii/S0378778804003391}

\bibitem{Kanda2005}
M.~Kanda, R.~Moriwaki, Y.~Kimoto,
  \href{http://link.springer.com/10.1007/s10546-004-5644-5}{{Temperature
  Profiles Within and Above an Urban Canopy}}, Boundary-Layer Meteorology
  115~(3) (2005) 499--506.
\newblock \href {http://dx.doi.org/10.1007/s10546-004-5644-5}
  {\path{doi:10.1007/s10546-004-5644-5}}.
\newline\urlprefix\url{http://link.springer.com/10.1007/s10546-004-5644-5}

\bibitem{Lemonsu2004}
A.~Lemonsu, C.~S.~B. Grimmond, V.~Masson,
  \href{http://journals.ametsoc.org/doi/full/10.1175/1520-0450(2004)043<0312:MTSEBO>2.0.CO;2}{{Modeling
  the Surface Energy Balance of the Core of an Old Mediterranean City:
  Marseille}}, Journal of Applied Meteorology 43~(2) (2004) 312--327.
\newblock \href
  {http://dx.doi.org/10.1175/1520-0450(2004)043<0312:MTSEBO>2.0.CO;2}
  {\path{doi:10.1175/1520-0450(2004)043<0312:MTSEBO>2.0.CO;2}}.
\newline\urlprefix\url{http://journals.ametsoc.org/doi/full/10.1175/1520-0450(2004)043<0312:MTSEBO>2.0.CO;2}

\bibitem{Gilboa2014}
I.~Gilboa, A.~Postlewaite, L.~Samuelson, D.~Schmeidler,
  \href{http://doi.wiley.com/10.1111/ecoj.12128}{{Economic Models as
  Analogies}}, The Economic Journal (2014) n/a--n/a\href
  {http://dx.doi.org/10.1111/ecoj.12128} {\path{doi:10.1111/ecoj.12128}}.
\newline\urlprefix\url{http://doi.wiley.com/10.1111/ecoj.12128}

\bibitem{Steemers2003}
K.~Steemers,
  \href{http://linkinghub.elsevier.com/retrieve/pii/S0378778802000750}{{Energy
  and the city: density, buildings and transport}}, Energy and Buildings 35~(1)
  (2003) 3--14.
\newblock \href {http://dx.doi.org/10.1016/S0378-7788(02)00075-0}
  {\path{doi:10.1016/S0378-7788(02)00075-0}}.
\newline\urlprefix\url{http://linkinghub.elsevier.com/retrieve/pii/S0378778802000750}

\end{thebibliography}







\end{document}